\documentclass[prx,twocolumn,superscriptaddress,amsmath,amssymb]{revtex4-1}
\usepackage{graphicx}
\usepackage{dcolumn}
\usepackage{bm}

\usepackage{amsmath}
\usepackage{amssymb}
\usepackage{amsthm}
\usepackage{pinlabel}
\usepackage{natbib}
\usepackage{epstopdf}
\usepackage{color}

\newcommand{\ua}{\uparrow}
\newcommand{\da}{\downarrow}
\newcommand{\no}{\nonumber}
\newcommand{\ph}{^{\phantom{\dagger}}}
\usepackage{braket}

\newcommand{\kk}{\mathbf{k}}

\newcommand{\sr}{\text{SrPt}_{3}\text{P}}
\newcommand{\ca}{\text{CaPt}_{3}\text{P}}
\newcommand{\la}{\text{LaPt}_{3}\text{P}}

\usepackage{braket}

\begin{document}

\title{Influence of electron-phonon coupling strength on signatures of even and odd-frequency superconductivity}  

\author{Alex Aperis}
\email{alex.aperis@physics.uu.se}
\affiliation{Department of Physics and Astronomy, Uppsala University, P.\,O. Box 516, SE-75120 Uppsala, Sweden}
\author{Eiaki V.~Morooka}
\affiliation{Department of Applied Physics, Aalto University, P.\,O. Box 11100, FI-00076, Aalto, Espoo, Finland}
\author{Peter M.~Oppeneer}
\affiliation{Department of Physics and Astronomy, Uppsala University, P.\,O. Box 516, SE-75120 Uppsala, Sweden}

\vskip 0.7cm
\date{\today}
\begin{abstract}
\noindent

The recently discovered APt$_3$P (A=Sr,Ca,La) family of superconductors offers a platform to study  frequency dependent superconducting phenomena as the electron-phonon coupling varies from weak to strong. Here we perform \textit{ab initio} Eliashberg theory calculations to investigate two such phenomena, the occurrence of dip-hump structures in the tunneling spectra and the magnetic field induced coexistence of even and odd frequency superconductivity in these compounds. By calculating the superfluid density, we make predictions for the occurrence of the paramagnetic Meissner effect as a hallmark of odd frequency pairing. Our results provide a link between two seemingly uncorrelated aspects of even and odd frequency superconductivity and provide theoretical guidance for the experimental identification of bulk odd frequency superconductivity in this material's family.
\end{abstract}

\date{\today}
\maketitle

\section{Introduction}

In his pioneering work on superconductivity \cite{eliashberg1960}, Eliashberg ingeniously combined the ground-breaking theories of Bardeen-Cooper-Schrieffer (BCS)  \cite{bardeen1957} and that of Migdal on the electron-phonon interaction in metals \cite{Migdal1958} into what is now established as the most successful theory for explaining  superconductivity in real materials \cite{Parks1969,Allen1983,carbotte1990,bennemann2008,Giustino2017}. 
Eliashberg theory generalizes the weak-coupling BCS description of superconductors by explicitly taking into account the retarded nature of the electron-boson interaction that mediates the Cooper pairing. As such, it constitutes the basis for a full microscopic description of the phenomenon from first principles \cite{Allen1983,Giustino2017}. Nowadays, the self-consistent solution of the Eliashberg equations, supplemented with \textit{ab initio} calculated input for electrons and phonons has evolved into an extremely powerful method for the materials' specific modeling of superconductors on the quantitative level \cite{Choi2002,Margine2013,Aperis2015,Sanna2018,Bekaert2019}. As new superconducting phenomena and materials continue to be discovered and old puzzles like the high-T$_c$ phenomenon  remain yet unanswered \cite{100years}, new developments in the Eliashberg theory of superconductivity and numerical solution of the respective Eliashberg equations remain at the focus of current research \cite{Grimaldi1995,Aperis2015,Rademaker2016,Aperis2018,Sanna2018,Schrodi2018,Gastiasoro2019,Dee,Schrodi2019}.

A profound manifestation of Eliashberg theory is that the superconducting  gap function acquires a frequency dependence due to the retarded nature of the pairing. Usually, the stronger the retardation, the stronger is the coupling between electrons and phonons (although not always \cite{Bohnen2001}). The more pronounced coupling introduces a significant frequency dependence in the superconducting phenomenology which can dramatically deviate from the weak-coupling predictions of BCS. Strong coupling, for example, can lead to the absence of a Hebel-Slichter peak in the nuclear spin relaxation rate of conventional superconductors \cite{Allen1991}. Perhaps the best known such effect is the depletion and subsequent enhancement of the quasiparticle density of states (DOS) that can be measured e.g.\ by scanning tunneling spectroscopy (STS) experiments  \cite{Scalapino1966,Parks1969}. 
This characteristic ``dip-hump'' shape is the smoking gun of strong coupling superconductivity \cite{Scalapino1966}. It explicitly relates to the frequency dependence of the gap function since it is produced by the competition between the real and imaginary parts of the frequency dependent superconducting condensate \cite{Scalapino1966}.

An as yet less explored aspect of the frequency dependence of the gap function concerns the intriguing possibility of having Cooper pairing with wavefunctions that are odd in time, i.e.\ odd frequency superconductivity \cite{Linder2017}. The possible existence of such a state was first postulated by Berezinskii on phenomenological grounds in his work on $^3$He \cite{Berezinskii1974new}. 
Since Cooper pairs in superconductors consist of electrons, the pair wavefunction has to be antisymmetric under particle exchange due to the Pauli exclusion principle. This property gives rise to the standard classification of the gap function by means of the symmetry under spatial and spin rotation  so that e.g.\ s-wave superconductors are spin singlet and p-wave are necessarily spin triplet, etc. \cite{Sigrist1991}. In the aforementioned classification, the frequency dependence of the gap function is either implicitly (BCS) or explicitly (Eliashberg) considered to be even. As Berezinskii pointed out, the antisymmetry of the pair amplitude may as well be fulfilled when it is odd in frequency; however now, due to the sign-change in the frequency sector, the orbital and spin quantum numbers of the pair are reshuffled \cite{Linder2017}. In such a case, new superconducting states can be envisioned like e.g.\ an odd-frequency s-wave, spin triplet (OST), also called Berezinskii state \cite{Berezinskii1974new} (which is the state that we will consider here). 

Clearly, this type of superconductivity is beyond the BCS picture but can naturally be described within Eliashberg theory \cite{Fuseya2003,Kusunose2011,Kusunose2011a,Matsumoto2012,Aperis2015}. Interestingly, Eliashberg theory investigations found that, in order to stabilize the pure OST against the prevalent even-frequency s-wave spin singlet (ESS) state requires a very retarded and strong coupling electron-phonon interaction \cite{Kusunose2011,Kusunose2011a}. This observation established the emergence of bulk OST pairing as yet another manifestation of strong coupling phenomena in superconductors.
The required retardation conditions were found to be so extreme that they hardly exist in real materials, and hence,  further ways to promote the OST state were proposed  \cite{Fuseya2003,Matsumoto2012}. As Matsumoto \textit{et al.}\ pointed out \cite{Matsumoto2012}, a plausible scenario is to break time-reversal symmetry in a usual ESS superconductor by applying an external magnetic field. In this case, by symmetry arguments (and as we shall see later, for Eliashberg theory calculations), an OST component is induced by the field and coexists with the ESS part \cite{Matsumoto2012,Aperis2015,Fukui2018,Fukui2019}. This approach indeed opens the possibility of observing odd frequency pairing in real materials as has been predicted by \textit{ab initio} Eliashberg theory calculations in the case of MgB$_2$ \cite{Aperis2015}.

Odd-frequency superconductivity has been previously proposed on several occasions \cite{Kirkpatrick1991,Balatsky1992,Coleman1994,Mazin2005,Linder2017} but has been mostly considered recently as a proximity induced effect in heterostructure interfaces \cite{Bergeret2005,Tanaka2007,Eschrig2008,Tanaka2012,Asano2014,Ebisu2015,Linder2015a}. 
Two main experimental signatures are associated with such a state; a zero bias peak in the tunneling spectra and a paramagnetic Meissner effect \cite{Linder2017}. Both have been recently reported to have been observed in heterostructures \cite{Pal2017,DiBernardo2015,Dibernardo2015b, Diesch2018}. However, the identification of odd frequency superconductivity in bulk materials remains elusive. The magnetic field induced coexistence of ESS and OST superconductivity could nonetheless provide a plausible means for experimental detection since temperature and magnetic field can both be used as tuning parameters in the lab. Yet, it has been shown that the OST component is only a small fraction of the dominant ESS and in realistic numerical simulations found to be as small as less than 1 meV \cite{Aperis2015}. This makes any attempt for experimental detection of OST more difficult. A way to overcome the  difficulties could be by adding more, relevant tuning parameters.

In this respect, the recently discovered \cite{Takayama} family of APt$_3$P (A = Sr, Ca, La) superconductors ($T_c = 8.4$, 6.6, and 1.5 K) offers the unique possibility to have the coupling and  retardation strength of the electron-phonon interaction as an additional tuning parameter in the quest for the discovery of bulk odd-frequency superconductivity.
These compounds are isostructural, i.e., they share the same tetragonal crystal structure with space group 4P/nmm \cite{Takayama}, yet they exhibit a large variation in their electron-phonon retardation profile and coupling strength $\lambda$, with $\lambda =$ 1.33, 0.86 and 0.57 for the compounds $\sr$, $\ca$ and $\la$, respectively \cite{Subedi2013}.  The computed variation of the coupling strength across the series is in good agreement with available specific heat jump measurements \cite{Takayama}. In addition, while LaPt$_3$P has one more electron per formula unit than SrPt$_3$P and CaPt$_3$P and therefore has a different band structure near the Fermi energy, SrPt$_3$P and CaPt$_3$P are isoelectronic compounds and therefore share a very similar electronic structure \cite{Nekrasov,Kang,Subedi2013}. These features make them an ideal testbed for exploring the influence of the coupling strength on the aforementioned frequency dependent superconducting phenomena in future experiments.

Here, taking as input the \textit{ab initio} calculated electron, phonon and electron-phonon properties of the APt$_3$P compounds, we numerically solve the Eliashberg equations for even and odd-frequency superconductivity in these materials. Starting with zero magnetic field properties, we accurately determine the STS spectra and predict how the dip-hump signatures vary across this family depending on the coupling strength. We subsequently perform calculations in the presence of magnetic fields and estimate the strength of the induced odd-frequency  superconductivity across these compounds. Further, we calculate the superfluid density in the coexistence phase of ESS and OST pairing varying temperature and magnetic field and predict signatures of the paramagnetic Meissner effect for these materials. We observe that for the strong coupling material, SrPt$_3$P, all effects are amplified while for the weak coupled, BCS-like compound LaPt$_3$P, no significant dip-hump or paramagnetic Meissner effect are to be expected. Interestingly, collecting all of our observations, we find an intimate relation among the dip-hump structure, the paramagnetic Meissner effect, and the fraction of the ESS Cooper pairs that become OST with the applied magnetic field. We provide several predictions for future experiments that could ultimately lead to the identification of odd-frequency bulk superconductivity in these materials.

The remaining of the paper is organized as follows: In Section II, we present in detail the theoretical framework that we employed to obtain our results, namely Eliashberg theory in Matsubara and real frequency space and its extension to include magnetic fields, odd-frequency pairing and the derivation of the superfluid density in this case. In Section III.A we present solutions of the standard Eliashberg equations for the APt$_3$P compounds and give a first discussion of standard phenomenology from weak to strong coupling. In Section III.B we present our real frequency Eliashberg solutions and our calculated tunneling spectra, focusing on dip-hump and related phenomena. In Section III.C we  present calculated solutions of the Eliashberg equations including the effect of an external Zeeman field that have odd-frequency superconductivity as self-consistent solution. In Section III.D we present our calculations for the paramagnetic Meissner effect and discuss the possible link of our results with a single strong coupling ratio. Section IV concludes this article with a short discussion and outlook.

\section{Methodology}

Our starting point for a microscopic description of electrons, phonons, electron-phonon and Zeeman interactions in a metal is the Hamiltonian,
\begin{eqnarray}\no
H&=&\sum_{{\bf k},\sigma}(\xi_{\bf k}+\sigma \mu_B h)c^\dagger_{{\bf k}\sigma}c\ph_{{\bf k}\sigma} 
+ \sum_{{\bf q},\nu}\hbar\omega_{{\bf q},\nu}\left(b^\dagger_{{\bf q}\nu}b\ph_{{\bf q}\nu}+\frac{1}{2}\right)\\\no
&+& \sum_{{\bf q},\nu}\sum_{{\bf k},\sigma}g^\nu_{\bf q}c^\dagger_{{\bf k+q}\sigma}c\ph_{{\bf k}\sigma}\left(b\ph_{{\bf q}\nu}+b^\dagger_{{\bf -q}\nu}\right) \\\label{h0}
&+& \frac{1}{2}\sum_{{\bf k,k',q}}\sum_{\sigma.\sigma'}c^\dagger_{{\bf k+q}\sigma}c^\dagger_{{\bf k'-q}\sigma'}V_{\bf q}c\ph_{{\bf k'}\sigma'}c\ph_{{\bf k}\sigma} ,
\end{eqnarray}
where ${\bf q}={\bf k'-k}$, $\sigma$ denotes the electron spin and $\nu$ indexes the different phonon branches. In the above $\xi_{\kk}$ is the electron energy dispersion, $h$ is an external magnetic field, $\omega\ph_{{\bf q},\nu}$ are branch-resolved phonon frequencies, $g^\nu_{{\bf q}}$ is the branch-resolved electron-phonon coupling vertex and $V_{{\bf q}}$ is the electron-electron Coulomb interaction. As usual, $c\ph_{\bf k}$ ($c^{\dagger}_{\bf k}$) and  $b\ph_{{\bf q}\nu}$ ($b^{\dagger}_{{\bf q}\nu}$) are the second quantized electron and phonon annihilation (creation) operators, respectively. It is worth mentioning that $\xi_{\kk}$, $\omega\ph_{{\bf q},\nu}$ and $g^\nu_{{\bf q}}$ are quantities that can nowadays be reliably calculated \textit{ab initio} with the use of Density Functional Theory (DFT) and Density Functional Perturbation Theory (DFTP) methods \cite{Giustino2017}. 

By introducing the Nambu spinor,
\begin{eqnarray}
\Psi_{\bf k}^\dagger=\frac{1}{\sqrt{2}}\left(c^\dagger_{{\bf k}\ua},c^\dagger_{{\bf k}\da},c\ph_{-{\bf k}\ua},c\ph_{-{\bf k}\da}\right),
\end{eqnarray}
that acts on the Pauli basis spanned by $\hat{\rho}_i\otimes\hat{\sigma}_j$, with $i,j=0,1,2,3$, Eq.\ (\ref{h0}) can be recast in the more compact form \cite{Aperis2015},
\begin{eqnarray}\no
H&=&\sum_{{\bf k}}\xi\ph_{\bf k}\Psi^\dagger_{{\bf k}}\hat{\rho}_3\hat{\sigma}_0 \Psi\ph_{{\bf k}} 
+\mu\ph_B h\sum_{{\bf k}}\Psi^\dagger_{{\bf k}}\hat{\rho}_3\hat{\sigma}_3 \Psi\ph_{{\bf k}}\\\no
&+& \sum_{{\bf q},\nu}\hbar\omega\ph_{{\bf q},\nu}\left(b^\dagger_{{\bf q}\nu}b\ph_{{\bf q}\nu}+\frac{1}{2}\right)\\\no
&+& \sum_{{\bf q},\nu}\sum_{{\bf kk'}}g^\nu_{\bf q}\Psi^\dagger_{{\bf k'}}\hat{\rho}_3\hat{\sigma}_0 \Psi\ph_{{\bf k}}\left(b\ph_{{\bf q}\nu}+b^\dagger_{{\bf -q}\nu}\right) \\\label{h012z}
&+& \frac{1}{2}\sum_{{\bf k,k',q}}\Psi^\dagger_{{\bf k'}}\hat{\rho}_3\hat{\sigma}_0\Psi\ph_{{\bf k'}}V\ph_{\bf q}\Psi^\dagger_{{\bf k}}\hat{\rho}_3\hat{\sigma}_0\Psi\ph_{{\bf k}}\,.
\end{eqnarray}
Rewriting the Hamiltonian in the so-called Nambu formalism \cite{nambu1960} the electronic Green's function becomes $4\times 4$ a matrix,
\begin{eqnarray}
\hat{G}({\bf k},\tau)=-\langle T_\tau \Psi\ph_{\bf k}(\tau)\otimes\Psi^\dagger_{\bf k}(0)\rangle ,
\end{eqnarray}
whose $2\times 2$ off-diagonal blocks have as elements anomalous propagators that describe scattering in the particle-particle channel and therefore superconductivity.
As usual, Eq.\ (\ref{h012z}) gives rise to the following coupled Dyson equations for the electron and phonon Green's functions, respectively,
\begin{eqnarray}\label{dyson1}
\hat{G}^{-1}({{\bf k},i\omega_n})=\hat{G}^{-1}_0({{\bf k},i\omega_n})-\hat{\Sigma}({{\bf k},i\omega_n}),\\\label{dyson2}
D^{-1}({{\bf q},iq_n})=D^{-1}_0({{\bf q},iq_n})-\Pi({{\bf q},iq_n}),
\end{eqnarray}
where $\hat{G}_{(0)}({{\bf k},i\omega_n})$ is the full (bare) electron Green's function, $D({{\bf q},iq_n})$ is the full and $D_0({\bf q},iq_n)=\sum_\nu D^\nu_0({\bf q},iq_n)=\sum_\nu\frac{-2\omega_{\bf q},\nu}{q^2_n+\omega^2_{{\bf q},\nu}}
$ is the bare phonon Green's function, respectively. 
The free matrix propagator has the form, 
\begin{eqnarray}\label{g02}
\hat{G}\ph_0({\bf k},i\omega_n)=\left(i\omega_n\hat{\rho}_0\hat{\sigma}_0-\xi\ph_{\bf k}\hat{\rho}_3\hat{\sigma}_0-\mu\ph_Bh\hat{\rho}_3\hat{\sigma}_3\right)^{-1}\,.
\end{eqnarray}
Here, $\omega_n=(2n+1)\pi T$, $q_n=2n\pi T$ are fermionic and bosonic Matsubara frequencies, respectively.
The electron and phonon self-energies are denoted as $\hat{\Sigma}({{\bf k},i\omega_n})$ and $\Pi({{\bf q},iq_n})$, respectively.
Adopting now Migdal's theorem \cite{Migdal1958}, vertex corrections to the self-energy can be neglected as long as $\omega_{ph}/\epsilon_F \ll 1$, where $\omega_{ph}$, $\epsilon_F$ are the characteristic phonon frequency and the Fermi energy, respectively. We also approximate the full phonon propagator as $D_0({\bf q},iq_n)$. This amounts to neglecting the feedback of the superconducting state on the phonons, which  in most cases should be small, but it includes the essential effects of the electron-phonon coupling in the normal state since $D_0({\bf q},iq_n)$ is calculated by first principles \cite{Allen1983}. 
The electron self-energy finally reads,
\begin{eqnarray}\no
&&\hat{\Sigma}({{\bf k},i\omega_n})=-T\sum_{{\bf k'},n'}\hat{\rho}_3\hat{\sigma}_0 \hat{G}({\bf k'},i\omega_{n'})\hat{\rho}_3\hat{\sigma}_0\\\label{s1}
&\times&\Bigl[\sum_\nu|g^\nu_{\bf q}|^2 D_\nu({\bf k-k'},i\omega_n-i\omega_{n'}) + V({\bf k-k'})\Bigl]\,,
\end{eqnarray}
where $\omega_n-\omega_{n'}=q_n$. The Coulomb interaction is included in Eq.\ (\ref{s1}) within the standard Hartree-Fock approximation which is on equal footing with the Migdal approximation for the electron-phonon term. It is assumed that the effect of the Coulomb interaction on the normal state has already been included in the calculation, therefore in Eq.\ (\ref{s1}) this term acts only on the paricle-particle off-diagonal parts of the matrix Green's function.

\subsection{Zero magnetic field case: standard Eliashberg equations\label{zeroeliashberg}}

From Eq.\ (\ref{dyson1}),(\ref{g02}),(\ref{s1}), we can obtain the standard Eliashberg equations by setting $h=0$ and taking the following \textit{Ansatz} for the self-energy:
\begin{eqnarray}\no
\hat{\Sigma}({\bf k},i\omega_n)&=&\left(1-Z({\bf k},i\omega_n)\right)i\omega_n\hat{\rho}_0\hat{\sigma}_0\\\label{s0}
&+&\chi({\bf k},i\omega_n)\hat{\rho}_3\hat{\sigma}_0+\phi_e\ph({\bf k},i\omega_n)\hat{\rho}_2\hat{\sigma}_2\,,
\end{eqnarray}
where $Z({\bf k},i\omega_n),\chi({\bf k},i\omega_n)$ are the respective mass and chemical potential renormalization functions and $\phi_e({\bf k},i\omega_n)$ is the superconducting pairing function in the spin-singlet, even frequency channel. For non-doped systems whose electronic Density of States (DOS) does not vary rapidly around the Fermi level, it is possible to integrate out the electron degrees of freedom away from the Fermi level and by doing so, one obtains $\chi({\bf k},i\omega_n)=0$ \cite{Allen1983}. As a final step, we will assume that all quantities of our theory are not significantly momentum dependent so that they can be considered isotropic, $Z({\bf k},i\omega_n)\approx Z(i\omega_n),\phi_e({\bf k},i\omega_n)\approx \phi_e(i\omega_n)$. Such an approximation is well justified for many superconductors where the electron-phonon interaction is almost isotropic \cite{carbotte1990} and this is the case for the APt$_3$P superconductors that are considered here \cite{Subedi2013}.

Taking all discussed approximations into account, the electron self-energy now has the simple form,
\begin{eqnarray}\label{s0}
\hat{\Sigma}(i\omega_n)=\left(1-Z(i\omega_n)\right)i\omega_n\hat{\rho}_0\hat{\sigma}_0+\phi_e\ph(i\omega_n)\hat{\rho}_2\hat{\sigma}_2\, ,
\end{eqnarray}
which, supplemented by the bare electron propagator, $\hat{G}_0({\bf k},i\omega_n)=(i\omega_n\hat{\rho}_0\hat{\sigma}_0-\xi\ph_{\bf k}\hat{\rho}_3\hat{\sigma}_0)^{-1}$ and the energy integrated Eq.\ (\ref{s1}) yields the celebrated system of two coupled Eliashberg equations \cite{eliashberg1960}:
\begin{widetext}
\begin{eqnarray}\label{El01iso}
Z(i\omega_n)&=&1+\frac{1}{2n+1}\sum_{n'}\lambda(\omega_n-\omega_{n'})\frac{\omega_{n'}}{\sqrt{\omega^2_{n'}+\Delta_e({i\omega_{n'}})^2}}\, , \\\label{El02iso}
Z(i\omega_n)\Delta_e(i\omega_n)&=&\pi T\sum_{n'}^{|\omega_{n'}|<\omega_c}\left\{\lambda(\omega_n-\omega_{n'})-\mu^*(\omega_c)\right\}\frac{\Delta_e(i\omega_{n'})}{\sqrt{\omega^2_{n'}+\Delta_e(i\omega_{n'})^2}}\,,
\end{eqnarray}
\end{widetext}
where $\Delta_e(i\omega_n)=\phi_e(i\omega_n)/Z(i\omega_n)$ is the superconducting gap function and
\begin{eqnarray}\label{lnn}
\lambda(\omega_n-\omega_{n'})=\int_0^\infty d\Omega \, \alpha^2F(\Omega) \frac{2\Omega}{(\omega_n-\omega_{n'})^2+\Omega^2} \, ,
\end{eqnarray}
is the electron-phonon coupling that depends on the Eliashberg function,
\begin{eqnarray}
\alpha^2F(\Omega)=N(0)\sum_{{\bf q}\nu} |g^\nu_{\bf q}|^2 \delta(\Omega-\omega\ph_{{\bf q}\nu})\,,
\end{eqnarray}
with $N(0)$ the DOS at the Fermi level. For the APt$_3$P family, the \textit{ab initio} calculated Eliashberg function turns out to be almost proportional to the phonon DOS, $F(\Omega)=\sum_{{\bf q}\nu}\delta(\Omega-\omega\ph_{{\bf q}\nu})$ \cite{Subedi2013}, thus indicating that the electron-phonon interaction is  to a good approximation isotropic ($|g^\nu_{\bf q}|^2\approx |g^\nu|^2$) in these compounds.
In Eq.\ (\ref{El02iso}) the Coulomb interaction has been renormalized to the so-called Coulomb pseudopotential $\mu^*(\omega_c)=0.1-0.2$ which is  taken as isotropic and is finite up to the cutoff energy $\omega_c$ \cite{Morel1962,Allen1983}. Remarkably, the value of the pseudopotential is the only free parameter in this formulation of the theory, since the electron DOS and the Eliashberg function can be provided by \textit{ab initio} calculations or experiment. In addition, \textit{ab initio} methods for the calculation of $\mu^*$ have been developed \cite{Sanna2018}.

\subsection{Calculation of tunneling spectra by means of self-consistent analytic continuation}

Equations (\ref{El01iso}) and (\ref{El02iso}) can be numerically solved by iteration until self-consistency is reached with a desired precision. The outcome provides the full Matsubara Green's function and can be used to calculate temperature dependent thermodynamic properties such as e.g.\ the critical temperature T$_c$. However, in order to obtain access to the real frequency dependence of the retarded Green's function, and therefore to spectroscopic properties such as the tunneling spectra, one has to perform a numerical analytic continuation to real frequencies. 
Here, we use the self-consistent analytic continuation method \cite{Marsiglio1988}. Being formally exact, this method does not suffer from any of the pathologies of the Pad\'e approximant method \cite{Vidberg1977} and can therefore provide the basis for extremely accurate calculations of tunneling spectra at any frequency.
The trade-off for such accuracy is that another system of mixed real and imaginary frequency equations needs to be solved self-consistently, using as input the results from Eqs.\ (\ref{El01iso}) and (\ref{El02iso}). These are the following equations,
\begin{widetext}
\begin{eqnarray}\label{ancont1}
\Delta(\omega)Z(\omega)&=&\pi T\sum_{n'}\left[\lambda(\omega-i\omega_{n'})-\mu^*(\omega_c)\right]\frac{\Delta(i\omega_{n'})}{\sqrt{R(i\omega_{n'})}}
+i\pi\int_{-\infty}^{\infty}d\omega' \Gamma(\omega,\omega')\alpha^2F(\omega')\frac{Z(\omega-\omega')\Delta(\omega-\omega')}{\sqrt{Z^2(\omega-\omega')R(\omega-\omega')}}, ~~~\\\label{ancont2}
Z(\omega)&=&1 + i\frac{\pi T}{\omega}\sum_{n'}\frac{\omega_{n'}}{\sqrt{R(i\omega_{n'})}}\lambda(\omega-i\omega_{n'})
+i\frac{\pi}{\omega}\int_{-\infty}^{\infty}d\omega' \Gamma(\omega,\omega')\alpha^2F(\omega')\frac{(\omega-\omega')Z(\omega-\omega')}{\sqrt{Z^2(\omega-\omega')R(\omega-\omega')}} ,
\end{eqnarray}
\end{widetext}
where $R(i\omega_n)=\omega_n^2+\Delta^2(i\omega_n)$, $R(\omega)=\omega^2-\Delta^2(\omega)$, $\Gamma(\omega,\omega')=\frac{1}{2}\left(\tanh{\frac{\omega-\omega'}{2T}}+\coth{\frac{\omega'}{2T}}\right)$ and $\lambda(\omega-i\omega_{n})=-\int_{-\infty}^{\infty}d\omega'\frac{\alpha^2F(\omega')}{\omega-i\omega_n-\omega'}$.
Once the real frequency gap function is calculated, we can compute the normalized superconducting DOS using the following expression
\begin{equation}\label{tun}
\frac{N_S(\omega)}{N(0)}=\text{Re} \frac{|\omega|}{\sqrt{\omega^2 - \Delta^2(\omega)}}\, ,
\end{equation}
which is directly proportional to the normalized differential conductance measured in STS experiments.

\subsection{Eliashberg theory for the coexistence of even and odd frequency superconductivity in the presence of external Zeeman fields\label{app1}}

We now return to the full Hamiltonian of Eq.\ (\ref{h0}) and consider a non-zero magnetic field.  As before, the electron self-energy is given by Eq.\ (\ref{s0}) and the full Green's function satisfies the respective Dyson equation of Eq.\ (\ref{dyson1}). We will follow the same procedure and set of approximations as outlined in detail in Sec.\ \ref{zeroeliashberg}. 
Since the application of a Zeeman magnetic field is expected to induce a tendency towards odd-frequency pairing \cite{Matsumoto2012}, the minimal matrix self-energy that we consider here has the form
\begin{eqnarray}\no
\hat{\Sigma}(i\omega_n)&=&\left(1-Z(i\omega_n)\right)i\omega_n\hat{\rho}_0\hat{\sigma}_0+\Sigma_h\ph(i\omega_n)\hat{\rho}_3\hat{\sigma}_3\\\label{s2}
&+&\phi_e\ph(i\omega_n)\hat{\rho}_2\hat{\sigma}_2
+i\phi_o\ph(i\omega_n)\hat{\rho}_1\hat{\sigma}_1,
\end{eqnarray}
where $Z(i\omega_n)$ and $\phi_e(i\omega_n)$ are the same quantities as defined before. However, due to the presence of the magnetic field there are two additional self-energy terms: $\Sigma_h\ph(i\omega_n)$ is the self-energy that renormalizes the external magnetic field and $i\phi_o\ph(i\omega_n)$ describes the possibility of realizing s-wave, odd-frequency spin triplet superconductivity \cite{Matsumoto2012}. Here we choose the magnetic field and the \textbf{d}-vector of the odd-frequency spin triplet superconductivity to lie along the $z$-axis.

From equations (\ref{s1}), (\ref{g02}) and (\ref{s2}) one can now derive the following system of four coupled self-consistent Eliashberg equations for the coexistence of even and odd frequency superconductivity in the presence of  Zeeman fields:
\begin{widetext}
\begin{eqnarray}\label{eqS1}
Z(i\omega_n)&=&1+\frac{\pi T}{2\omega_n}\sum_{n',\pm}\lambda(\omega_n-\omega_{n'})\frac{\omega_{n'}\pm i \tilde{\textrm{H}}(i\omega_{n'})}{\Bigl[ \left(\omega_{n'}\pm i\tilde{\textrm{H}}(i\omega_{n'})\right)^2+\left(-\Delta_e\ph(i\omega_{n'})\pm i\Delta_o\ph(i\omega_{n'})\right)^2\Bigl]^\frac{1}{2}} \, ,
\\\label{eqS2}
\Sigma_h\ph(i\omega_n)&=&\frac{\pi T}{2}\sum_{n',\pm}\lambda(\omega_n-\omega_{n'})\frac{\tilde{\textrm{H}}(i\omega_{n'})\mp i\omega_{n'}}{\Bigl[ \left(\omega_{n'}\pm i\tilde{\textrm{H}}(i\omega_{n'})\right)^2+\left(-\Delta_e\ph(i\omega_{n'})\pm i\Delta_o\ph(i\omega_{n'})\right)^2\Bigl]^\frac{1}{2}} \, ,
\\\label{eqS3}
Z(i\omega_n)\Delta_e\ph(i\omega_n)&=&\frac{\pi T}{2}\sum_{n',\pm}\left[\lambda(\omega_n-\omega_{n'})-\mu^*(\omega_c)\right]\frac{\Delta_e\ph(i\omega_{n'})\mp i\Delta_o\ph(i\omega_{n'})}{\Bigl[ \left(\omega_{n'}\pm i\tilde{\textrm{H}}(i\omega_{n'})\right)^2+\left(-\Delta_e\ph(i\omega_{n'})\pm i\Delta_o\ph(i\omega_{n'})\right)^2\Bigl]^\frac{1}{2}} \, , 
\\\label{eqS4}
Z(i\omega_n)\Delta_o\ph(i\omega_n)&=&\frac{\pi T}{2}\sum_{n',\pm}\left[\lambda(\omega_n-\omega_{n'})-\mu^*(\omega_c)\right]
\frac{\Delta_o\ph(i\omega_{n'})\pm i \Delta_e\ph(i\omega_{n'})}{\Bigl[ \left(\omega_{n'}\pm i\tilde{\textrm{H}}(i\omega_{n'})\right)^2+\left(-\Delta_e\ph(i\omega_{n'})\pm i\Delta_o\ph(i\omega_{n'})\right)^2\Bigl]^\frac{1}{2}}\,. ~~~
\end{eqnarray}
\end{widetext}
In the above, $\textrm{H}(i\omega_{n})=\Sigma_h\ph(i\omega_{n})+\mu\ph_B h$, $\tilde{\textrm{H}}(i\omega_{n})=\textrm{H}(i\omega_{n})/Z(i\omega_{n})$ and $\Delta_o(i\omega_{n})=\phi_o(i\omega_{n})/Z(i\omega_{n})$ in complete analogy with its even frequency counterpart.

By inspection of Eqs.\ (\ref{eqS1})-(\ref{eqS4}), one can observe that these have a structure that is markedly different from those of Eqs.\ (\ref{El01iso}) and (\ref{El02iso}). For example, in contrast to Eq.\ (\ref{El02iso}) where the right-hand-side (rhs) is proportional only to $\Delta_e(i\omega_{n'})$, the rhs of Eq.\ (\ref{eqS3}) contains two terms. Setting $\Delta_e(i\omega_{n'})=0$ and $\Delta_o(i\omega_{n'})=0$ on the rhs of Eqs.\ (\ref{eqS3}) and Eq.\ (\ref{eqS4}), respectively leads to equations of the form
\begin{widetext}
\begin{eqnarray}\label{eqS3b}
Z(i\omega_n)\Delta_e\ph(i\omega_n)&=&\frac{\pi T}{2}\sum_{n',\pm}\left[\lambda(\omega_n-\omega_{n'})-\mu^*(\omega_c)\right]\frac{\mp i\Delta_o\ph(i\omega_{n'})}{\Bigl[ \left(\omega_{n'}\pm i\tilde{\textrm{H}}(i\omega_{n'})\right)^2-\Delta_o\ph(i\omega_{n'})^2\Bigl]^\frac{1}{2}} \, ,
\\\label{eqS4b}
Z(i\omega_n)\Delta_o\ph(i\omega_n)&=&\frac{\pi T}{2}\sum_{n',\pm}\left[\lambda(\omega_n-\omega_{n'})-\mu^*(\omega_c)\right]
\frac{\pm i \Delta_e\ph(i\omega_{n'})}{\Bigl[ \left(\omega_{n'}\pm i\tilde{\textrm{H}}(i\omega_{n'})\right)^2+\Delta_e\ph(i\omega_{n'})^2\Bigl]^\frac{1}{2}}\, ,
\end{eqnarray}
\end{widetext}
where the left-hand-side gap functions are to be understood as induced due to the rhs counterparts being non-zero. Setting now the effective magnetic field $\tilde{H}(i\omega_{n'})$ to zero on the rhs of above  yields exactly zero after the $\pm$-sum is performed. Thus, it becomes evident that even and odd frequency solutions can be induced and therefore coexist once a magnetic field is turned on. We shall see below that this is indeed the case in our numerical solutions of Eqs.\ (\ref{eqS1})-(\ref{eqS4}). A similar mechanism for inducing the coexistence of otherwise competing order parameters with an applied magnetic field has been discussed before in the context of BCS theory \cite{Aperis2008,Aperis2010}.


\subsection{Derivation of the Magnetic Penetration Depth}

We complete our Methodology section with the derivation for the expression of the magnetic penetration depth in the case of coexisting even and odd-frequency superconductivity under external Zeeman fields.

The magnetic penetration depth, $\lambda$, is related to the superfluid density of the superconductor, $\rho_s$, via $\rho_s\propto\lambda^{-2}$ whereas $\rho_s$ is itself proportional to the so-called London kernel 
that describes the local, static current response to an applied transverse vector potential, $J_\alpha=-Q_{\alpha\beta}A^\beta$, with $\alpha,\beta=x,y$ \cite{Schrieffer2018}. 
Therefore, the normalized aforementioned quantities are related to each other by the following equations,
\begin{eqnarray}\label{superfluid0}
\frac{\lambda^{-2}(T,h)}{\lambda^{-2}(0,h)}=\frac{Q(T,h)}{Q(0,0)}=\frac{\rho_s(T,h)}{\rho_s(0,0)}\,.
\end{eqnarray}
Since $Q(T,h)$ is a static response function, it can be calculated by taking the double derivative with respect to the vector potential of the system's free energy \cite{Peotta2015}:
\begin{equation}\label{london0}
Q_{\alpha}^{\phantom{a}\beta} = - \frac{\delta^2 F}{\delta A^{\alpha} \delta A_{\beta}} \, ,
\end{equation}
where $F$ is the free energy that contains the effect of the magnetic field coupling to the electron orbital motion. The effect of the vector potential can be incorporated using the Peierls  substitution $\kk \rightarrow \kk - q \bf{A}$, so that the bare Green's function becomes modified as:
\begin{eqnarray}\no
\hat{G}^{-1}_0({\bf k},i\omega_n)=i\omega_n \hat{\rho}_0\hat{\sigma}_0-\xi({\bf k}-q{\bf A}\hat{\rho}_3\hat{\sigma}_0)\hat{\rho}_3\hat{\sigma}_0 \\\label{bareG2}
- h\hat{\rho}_3\hat{\sigma}_3 \, ,
\end{eqnarray} 
and the relevant expression for the superconducting free energy has the form,
\begin{eqnarray}\label{freen}
F_S&=&-\frac{T}{2}\sum_{\textbf{k},n}\text{Tr}\left\{ \ln{[-\hat{G}^{-1}(\textbf{k}-q\textbf{A},i\omega_n)}]\right\}\,.
\end{eqnarray}
Using then Eqs.\ (\ref{dyson1}), (\ref{s2}), (\ref{bareG2}) and ({\ref{freen}}) in Eq.\ (\ref{london0}) yields,
\begin{eqnarray}\no
Q_{\alpha\beta}&=&\frac{e^2}{2}T\sum_{{\bf k},n}\Bigl(\nabla_\alpha\xi_{{\bf k}}\nabla_\beta\xi_{{\bf k}}\text{Tr}\left\{\hat{G}({\bf k},i\omega_n)\hat{G}({\bf k},i\omega_n)\right\}\\
&+&\nabla^2_{\alpha\beta}\xi_{{\bf k}}\text{Tr}\left\{\hat{\rho}_3\hat{G}({\bf k},i\omega_n)\right\}\Bigl) .
\end{eqnarray}
Assuming an isotropic medium and turning the ${\bf k}$-sum into an energy integral, the second term of the above expression vanishes and we are left with:
\begin{eqnarray}
Q&=&\frac{e^2}{2}T\sum_{i\omega_n}N_0 \upsilon_F^2\int_{-\infty}^\infty d\xi\text{Tr}\left\{\hat{G}(\xi,i\omega_n)\hat{G}(\xi,i\omega_n)\right\} ,
\end{eqnarray}
where $\upsilon_F$ is the Fermi velocity. After performing the integral and rearranging the terms we arrive at the final expression,
\begin{widetext}
\begin{eqnarray}\label{londonfinal}
Q(T,h)&=&e^2 \upsilon^2_F N(0)\pi T\sum_{n} \sum_\pm \frac{\left(\phi_e(i\omega_n)\pm i\phi_o(i\omega_n)\right)^2}{\left[-\left(H(i\omega_n)\pm i\omega_n Z(i\omega_n)\right)^2+\left(\phi_e(i\omega_n)\pm i\phi_o(i\omega_n)\right)^2\right]^{\frac{3}{2}}} .
\end{eqnarray}
\end{widetext}
One can show that the above quantity is identically real  and it scales as $\propto(\phi_e(i\omega_n)^2-\phi_o(i\omega_n)^2)$, so that the odd-frequency term has a paramagnetic contribution. When the odd-frequency term dominates over the even frequency one, the London kernel and therefore the superfluid density becomes negative and the superconductor no longer exhibits diamagnetic properties. In other words, according to $\lambda\propto n_S^{-1/2}$, the penetration depth becomes purely imaginary and magnetic fields are no longer screened by the superconductor.
Lastly, 
as a crosscheck, it is straightforward to show that setting $h=\Sigma_h(i\omega_n)=\phi_o(i\omega_n)=0$ in Eq.\ (\ref{londonfinal}) gives the standard isotropic Eliashberg result \cite{Marsiglio1990, Golubov2002}.

\subsection{Details of numerical solution of the Eliashberg equations \label{app3}}

The sets of coupled self-consistent equations (\ref{El01iso})-(\ref{El02iso}), (\ref{ancont1})-(\ref{ancont2}) and (\ref{eqS1})-(\ref{eqS4}), were solved numerically supplemented with the \textit{ab initio} calculated electron, phonon and electron-phonon coupling \cite{Subedi2013}. In order to ensure a good accuracy for both the even and the odd-frequency gap functions, we imposed a strict convergence criterion of $\frac{x_n-x_{n-1}}{x_n}<10^{-9}$ and used up to $5\times 10^4$ Matsubara frequencies. In all calculations we used a cut-off frequency $\omega_c=10\times\omega_{ln}$ for the Coulomb pseudopotential, where the logarithmic frequency,
\begin{eqnarray}\label{wln}
\omega_{ln}=\exp\left[\frac{2}{\lambda}\int_0^\infty d\Omega \, \ln{(\Omega)}\alpha^2F(\Omega) /\Omega\right] ,
\end{eqnarray}
provides an estimate of the characteristic phonon energy scale \cite{Allen1975}. The value of $\mu^*$ was determined so that the experimental T$_c$ is accurately captured.

\section{Results and Discussion}

\subsection{Zero magnetic field solution of the Eliashberg equations  on the imaginary axis}

In this section we present our zero magnetic field numerical solution of the Eliashberg equations using as input the  DFT and DFPT data reported in Ref.\ \cite{Subedi2013}. Essentially, the calculations in this section are a repetition of those in Ref.\ \cite{Subedi2013}. However, it is worth presenting them first because we use a different code to solve the Eliashberg equations and therefore it is instructive to compare the results; second, the accompanying discussion will help elucidate the role of strong-coupling in the superconducting properties of the family series. Employing the same \textit{ab initio} input new results will be presented in the subsequent sections.

\begin{figure}[h]
\begin{center}
  \includegraphics[width=0.95\linewidth]{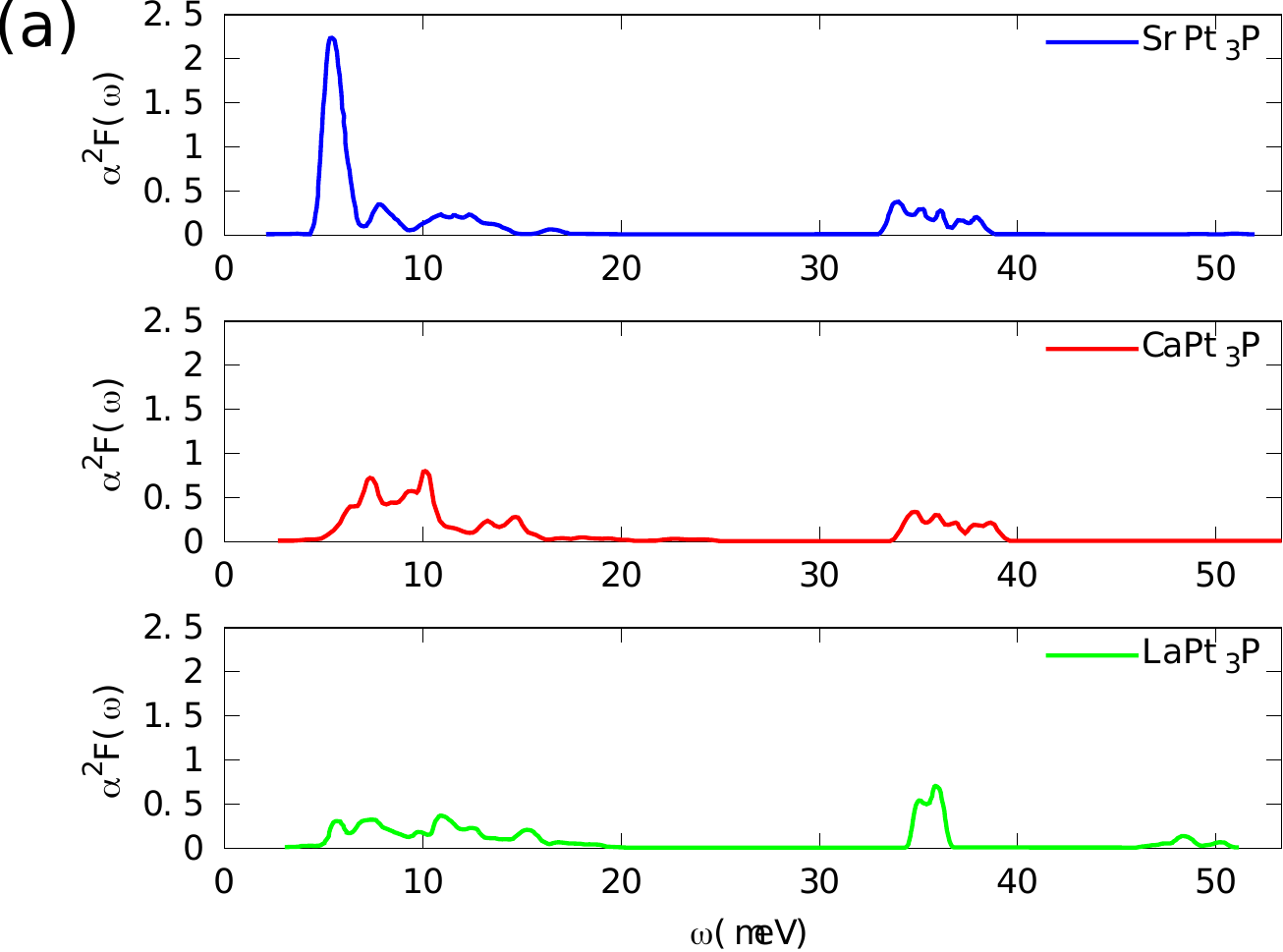}
    \includegraphics[width=0.8\linewidth]{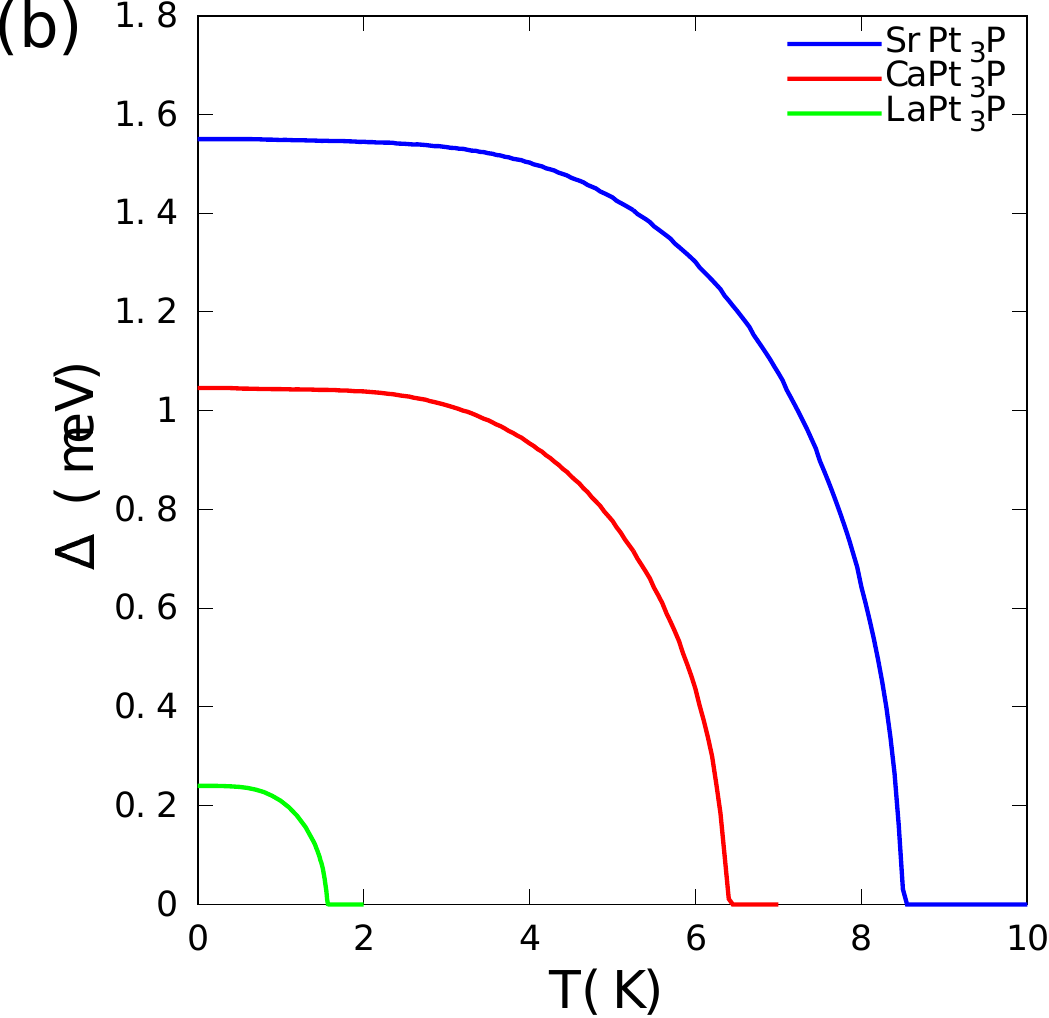}
  \caption{Top: The \textit{ab initio} computed Eliashberg functions $\alpha^{2}F(\omega)$ for SrPt$_{3}$P, CaPt$_{3}$P and LaPt$_{3}$P, respectively \cite{Subedi2013}. The area of the functions (weighted by 1/$\omega$) corresponds to the coupling strength so that $\sr$ is strongly coupled with $\lambda = 1.33$ due to strong retardation effects, while $\la$ is in the weak-coupling BCS regime with $\lambda = 0.57$. Bottom: Calculated energy gap, $\Delta(\omega\approx 0,T)$ in meV for SrPt$_{3}$P, CaPt$_{3}$P and LaPt$_{3}$P. Observe the size differences with the gap of $\la$ being an order of magnitude smaller than the other two.}\label{fig1}
  \end{center}
\end{figure}

The calculated Eliashberg functions for $\sr,  \ca$ and $\la$  \cite{Subedi2013} are shown below in Figure \ref{fig1}(a). In practice, these are the only input data needed to calculate all other relevant quantities. 
The electron-phonon coupling constant, $\lambda$, can easily be calculated by setting $\omega_n=\omega_{n'}$ in Eq.\ (\ref{lnn}), 
$\lambda=2\int_0^\infty d\Omega \, \alpha^2F(\Omega) /\Omega$,
whereas $\omega_{ln}$ is calculated from Eq.\ (\ref{wln}). Inserting each Eliashberg function in Eq.\ (\ref{lnn}) and a respective chosen value of $\mu^*$, we solved Eqs.\ (\ref{El01iso}) and (\ref{El02iso}) for different temperatures. The results for $\Delta(i\omega_{n=0})$ are shown in Fig.\ \ref{fig1}(b). 
The values obtained for various quantities are given in Table \ref{table1}. The first two columns are in exact agreement with Ref.\ \cite{Subedi2013} (as they should), while the rest of the columns differ very slightly except for the value of the Coulomb pseudopotential for LaPt$_3$P for which we find $\mu^*=0.16$ instead of $\mu^*=0.11$ of Ref.\ \cite{Subedi2013}. 
\begin{ruledtabular}
\begin{table}[ht]
 \centering
 \caption{Given are, for each material $\sr$, $\ca$, and $\la$, the coupling strength $\lambda$, the logarithmic phonon frequency $\omega_{\text{ln}}$, the  transition temperature $T_{c}$, the ratios $2\Delta(0)/T_{c}$ and $T_{c}/\omega_{ln}$ and the effective Coulomb pseudopotential $\mu^{*}$.  The values in parentheses are the experimental $T_c$ values which are taken from Ref.\ \cite{Takayama},  the other values are our calculated values based on \textit{ab initio} input from Ref.\ \cite{Subedi2013}.}
 \begin{tabular}{c |c c c c c c c}
  $\phantom{a}$&$\lambda$& $\omega_{\text{ln}}$ (K) & $T_{c}$ (K) & $2\Delta(0)/T_{c}$ & $T_{c}/\omega_{\ln}$ & $\mu^{*}$\\
 \hline 
 SrPt$_{3}$P & 1.33 & 77 & 8.56 (8.4) & 4.24  & 0.111 &0.11\\
 CaPt$_{3}$P & 0.85 & 110 & 6.45 (6.6) & 3.79  & 0.058 &0.11\\
 LaPt$_{3}$P & 0.57 & 118 & 1.58 (1.5) & 3.59  & 0.013 &0.16\\
 \end{tabular}\label{table1}
\end{table}
\end{ruledtabular}
Note that $\ca$ and $\sr$ are isoelectronic, and therefore an equal Coulomb potential $\mu^*$ can be expected for these compounds. $\la$ has a different electronic structure, since La contributes one more electron to the electronic bands. 
In the same Table we show the experimental $T_c$'s \cite{Takayama} in parentheses next to the calculated ones. The agreement between experiment and theory is extremely good. The accuracy of Eliashberg theory for the superconducting quantities is also corroborated by the agreement between the calculated  and measured specific heat jump at $T_c$ \cite{Subedi2013,Takayama} (not shown here).

From Fig.\ \ref{fig1}(a) we observe that for all three compounds the coupling spectral  weight is distributed roughly between $5-15$ meV and $35-40$ meV, and that their Eliashberg functions look quite similar in this respect. However, as we move from $\la$ to $\sr$, the spectral weight at low energies develops a pronounced peak which stems from the softening of the Pt in-plane breathing mode \cite{Subedi2013}. This effect is responsible for the variation of the coupling constant across the family since the pronounced low-energy spectrum leads to a higher $\lambda$. In addition, the logarithmic frequency is gradually pushed to lower energies, as well, indicating the strongly retarded nature of the electron-phonon interaction in $\sr$. Looking at Table \ref{table1}, this results in $\sr$ having a large coupling constant, $\lambda=1.33$ that yields a relatively high $T_c$, $\la$ having a much lower coupling and $T_c$ and $\ca$ being somewhere in the middle. The  gap over $T_c$ ratios are $2\Delta(0)/T_{c} = 4.24$,  3.76, and 3.59, respectively, for $\sr$, $\ca$ and $\la$, placing $\sr$ well in the strong coupling regime while $\la$ lies well in the weak-coupling BCS regime ($2\Delta(0)/T_{c}=3.53$ for BCS).

From the above, it is evident that the APt$_3$P family of compounds presents a rare example of materials where the superconducting phenomenology not only varies from weak to strong coupling but also in a simple, intuitive way by increased retardation. This makes them ideal for studying the evolution of phenomena that depend on retardation and are therefore absent in weakly-coupled (BCS) superconductors, such as the dip-hump effect in the tunneling spectra and the occurrence of odd-frequency superconductivity as we will show below.

\subsection{Real frequency Eliashberg solutions and tunneling spectroscopy for APt$_3$P (A = Sr, Ca, La)}

As discussed above, evidence of strong coupling effects across the phosphide family can already be obtained by considering  the gap over $T_c$. Here, we make a decisive step forward by studying how the differential conductance behaves for these materials within Eliashberg theory. It is known that this quantity can deviate greatly from the one calculated within BCS theory for frequencies higher than the gap-edge. Specifically, the stronger the coupling is, the more structure $dI/dV$ has as a function of frequency for $\omega>\Delta$ \cite{Scalapino1966}. 

\begin{figure*}[th!]
\centering
\includegraphics[width=0.99\linewidth]{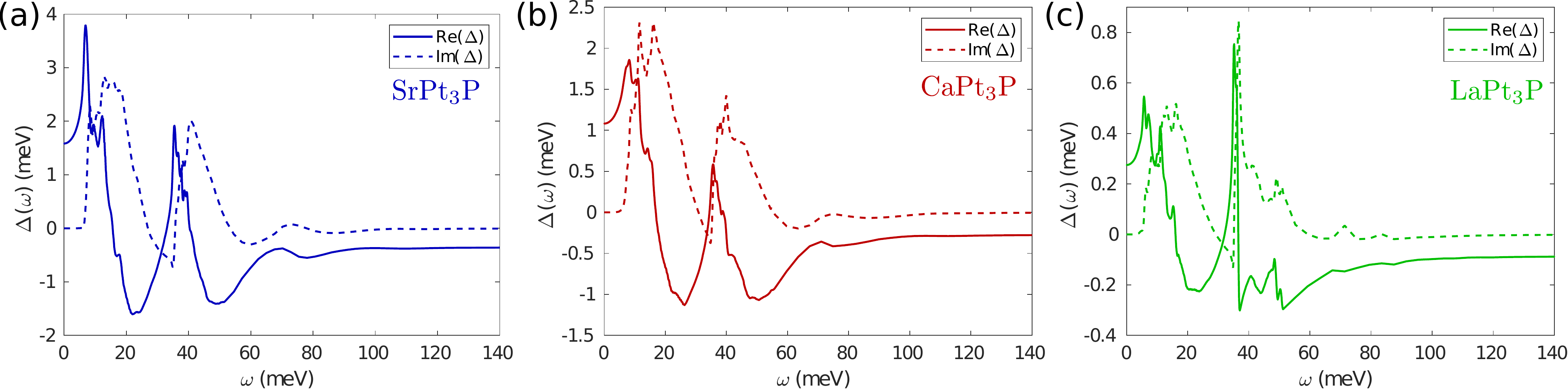}
\caption{Calculated real (solid) and imaginary (dotted) $\Delta(\omega)$ for (a) $\sr$ at $T$ = 1 K (b) $\ca$ at $T$ = 1 K and (c) $\la$ at $T$ = 0.3 K.} 
\label{fig2}
\end{figure*}

To obtain the tunneling spectra we first analytically continue the results of the previous section following the self-consistent analytic continuation method \cite{Marsiglio1988}. We do so by inserting the Matsubara space solutions of Eqs.\ (\ref{El01iso}) and (\ref{El02iso}) into Eqs.\ (\ref{ancont1}) and (\ref{ancont2}) and performing another self-consistent calculation. The real part of $\Delta(\omega)$ is related to the condensation energy of the Cooper pairing while the imaginary part is a measure of damping of the Cooper pairs \cite{Scalapino1966,Zeitschrift}. The calculated real and imaginary part of $\Delta(\omega)$ for $\sr$, $\ca$ and $\la$ at low temperatures are shown in Fig.\ \ref{fig2}.

\begin{figure}[t!] 
\label{chdb}
\includegraphics[width=0.8\linewidth]{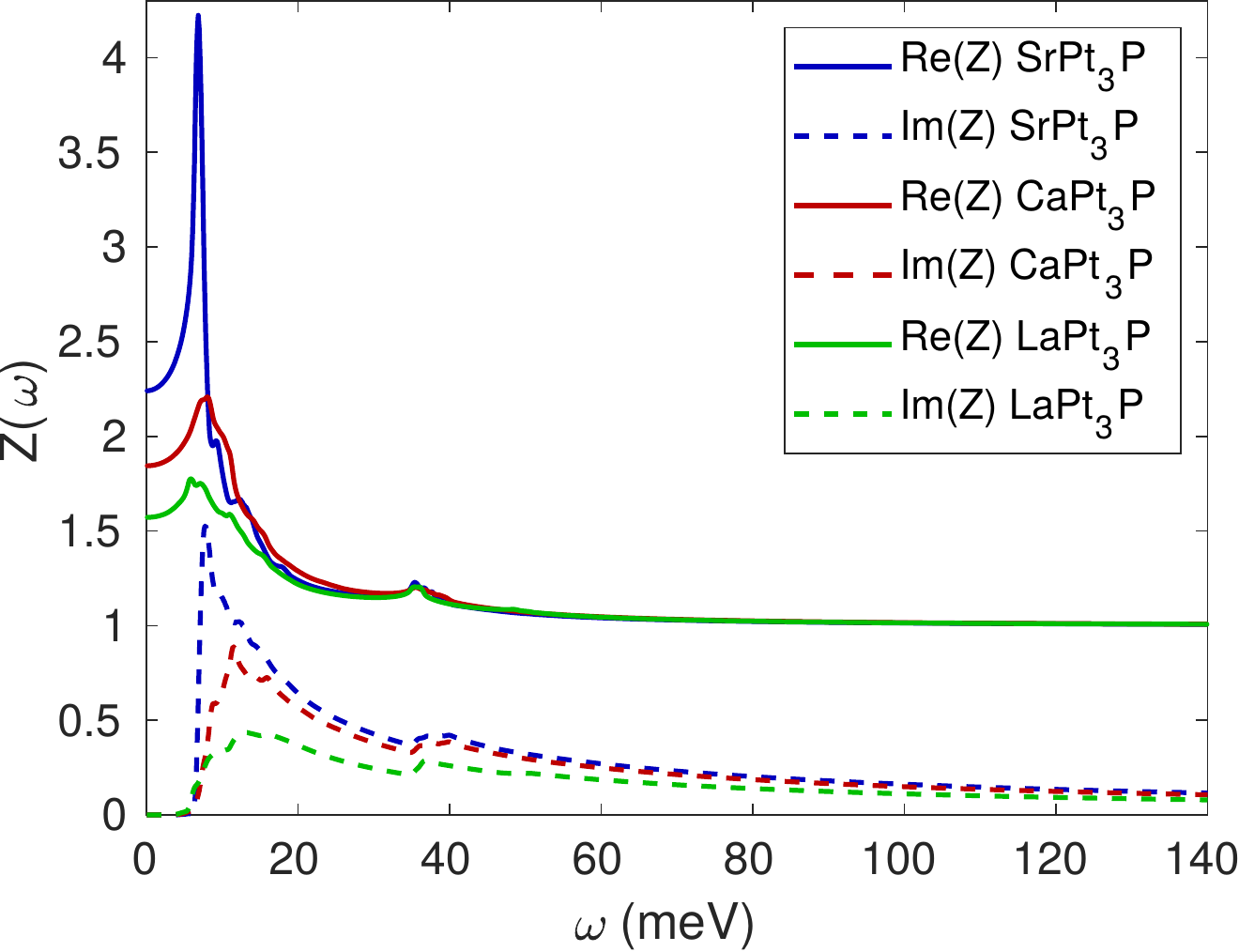}
\caption{Calculated real (solid) and imaginary part (dashed line) of $Z(\omega)$ for $\sr$ at $T$ = 1 K (blue lines), $\ca$ at $T$ = 1 K (red lines), and $\la$ at $T$ = 0.3 K (green lines).} 
\label{fig3}
\end{figure}

There are two well separated regions where $\Delta(\omega)$ peaks for all three compounds. These  correspond to the energies where the respective Eliashberg functions peak as shown in  Fig.\ \ref{fig1}(a). For example, the sharp peak near 7 meV in the Eliashberg function of $\sr$ is reflected in the respective $\rm{Re}\,\Delta(\omega)$ at the same energy. For completeness, we also show the real and imaginary parts of $Z(\omega)$ in Fig.\ \ref{fig3} which also exhibits a two peak structure that reflects those of the Eliashberg functions. The strong coupling to the low lying 7 meV phonon mode in $\sr$ is clearly seen and contrasted with the ones for $\ca$ and $\la$ where the respective peaks are much weaker. 

\begin{figure*}[bth!]
\includegraphics[width=0.99\linewidth]{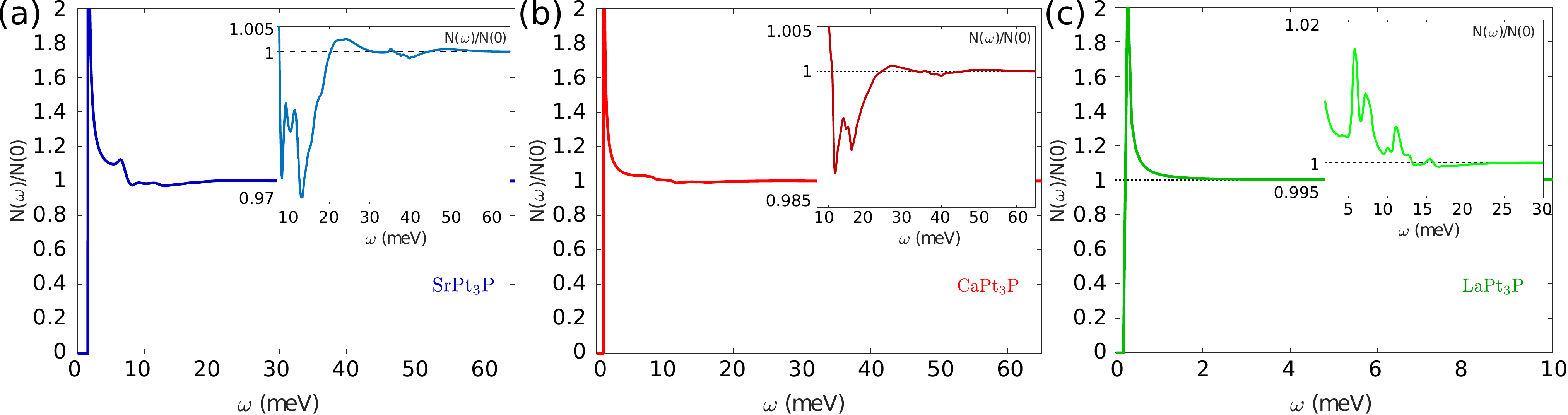}
\caption{Calculated normalized superconducting DOS for (a) $\sr$ at $T$ = 1 K (blue) (b) $\ca$ at $T$ = 1 K (red) and (c) $\la$ at $T$ = 0.3 K (green). Each figure includes a panel with a zoom in the dip-hump region. The dip-hump signature varies as we move across the family members depending on the coupling strength.}\label{fig4}
\end{figure*}

With the real frequency Eliashberg results, we proceed to calculate the normalized tunneling spectra from Eq.\ (\ref{tun}). The calculated results are shown in Fig.\ \ref{fig4} and correspond to the input shown in Fig.\ \ref{fig3} for each family member.

As seen in in Fig.\ \ref{fig3}, at the gap-edge, where $\Delta(\omega) = \omega$, the superconducting density of states diverges while for smaller frequencies the DOS is zero since we have a plain s-wave superconducting gap. As we move from $\sr$ to $\la$, the DOS gradually looses any visible structure and resembles closely a perfect BCS curve. This is again a clear manifestation of strong to weak coupling phenomenology in these materials. For energies roughly equal to $\omega_{ln}+\Delta_g$, with $\Delta_g$ the gap edge, the DOS of all three compounds exhibits a hump which is related to the dominant phonon mode \cite{Scalapino1966}. This hump almost disappears with decreasing the coupling, as well.

For energies higher than the characteristic phonon energy scale, $\sr$ and $\ca$ exhibit dip-hump structures between $10-23$ meV and $14-27$ meV, respectively (see also insets of Fig.\ \ref{fig4}). Such dip-hump signatures are the smoking gun of strongly coupled superconductivity \cite{Scalapino1966,Zeitschrift}. For $\la$, there exists a tiny dip-hump between $14-25$\,meV whose magnitude is negligible so that it is not discernible in the inset of Fig.\ \ref{fig4}(c) unless we zoom in very closely. The mechanism behind the dip-hump phenomenon is a competition between the real and imaginary parts of $\Delta(\omega)$, as can be seen by performing an expansion of Eq.\ (\ref{tun}) for $\Delta(\omega)/\omega \ll1$ \cite{Scalapino1966,Varelogiannis1995}:
\begin{equation}
\! \! \frac{N(\omega)}{N(0)}\!\approx \! 1+\frac{1}{2}\!\left[\left(\frac{\rm{Re}\left\{\Delta(\omega)\right\}}{\omega}\right)^2\!-\! \left(\frac{\rm{Im}\left\{\Delta(\omega)\right\}}{\omega}\right)^2\right]\,.~
\end{equation}
From Figs.\ \ref{fig2}(a),(b) one can see that this is indeed the physical picture. Interestingly, the distance between the energies where the dip and the hump occurs is predicted to be the same for $\sr$ and $\ca$ despite the fact that their coupling strength is markedly different. This somewhat appears to contrast previous findings where this distance was associated to the coupling strength \cite{Varelogiannis1995,Aperis2018}. However, in those works, the characteristic phonon frequency was kept constant while varying the interaction strength. Here, $\omega_{ln}$ decreases as the coupling increases so that the dip-hump energy distance stays the same. Therefore, in the case of the $\sr$ and $\ca$, the difference in coupling strength can be detected in tunneling experiments as a relative difference in the height of the dips and humps. For the strongly coupled $\sr$, a second dip hump may also be in principle observable as can be seen by the inset of Fig.\ \ref{fig4}(a). 

\subsection{Magnetic field induced coexistence of even and odd frequency superconductivity}

\begin{figure}[th!]
\centering
\includegraphics[width=0.490\linewidth]{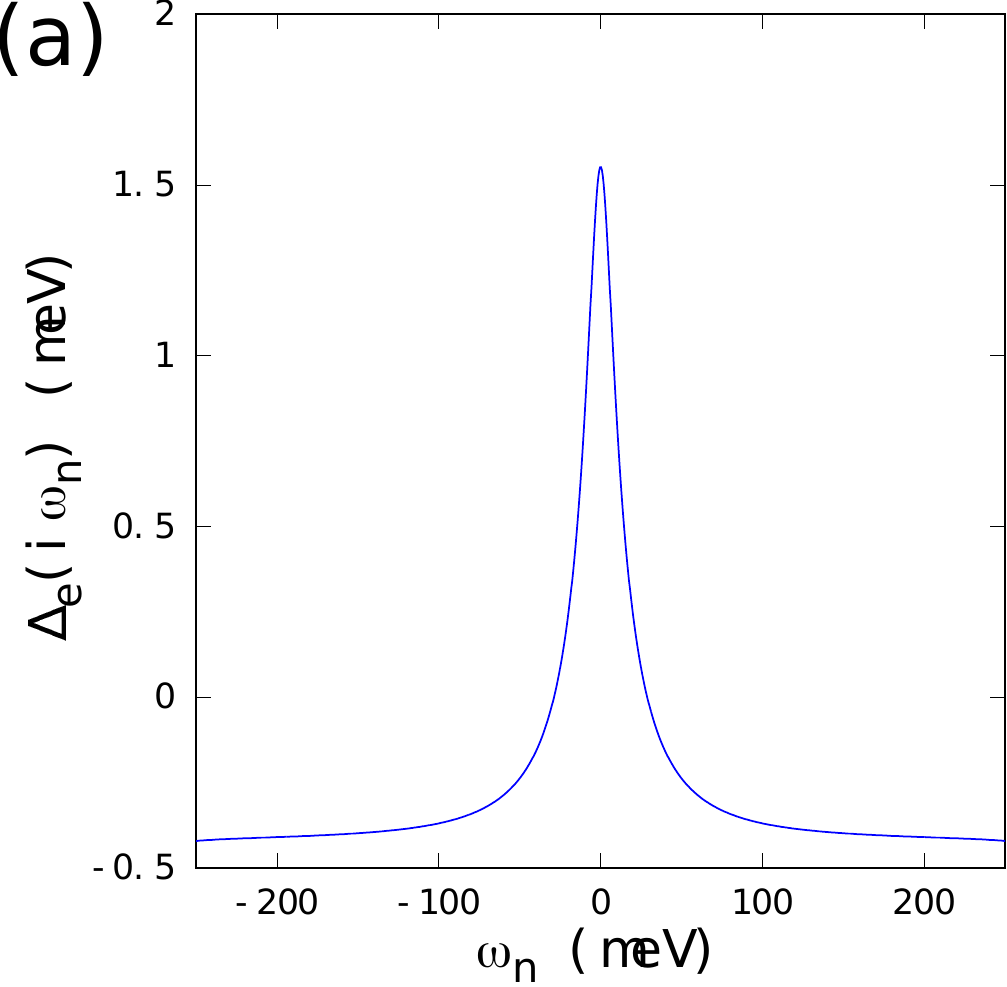}
\includegraphics[width=0.490\linewidth]{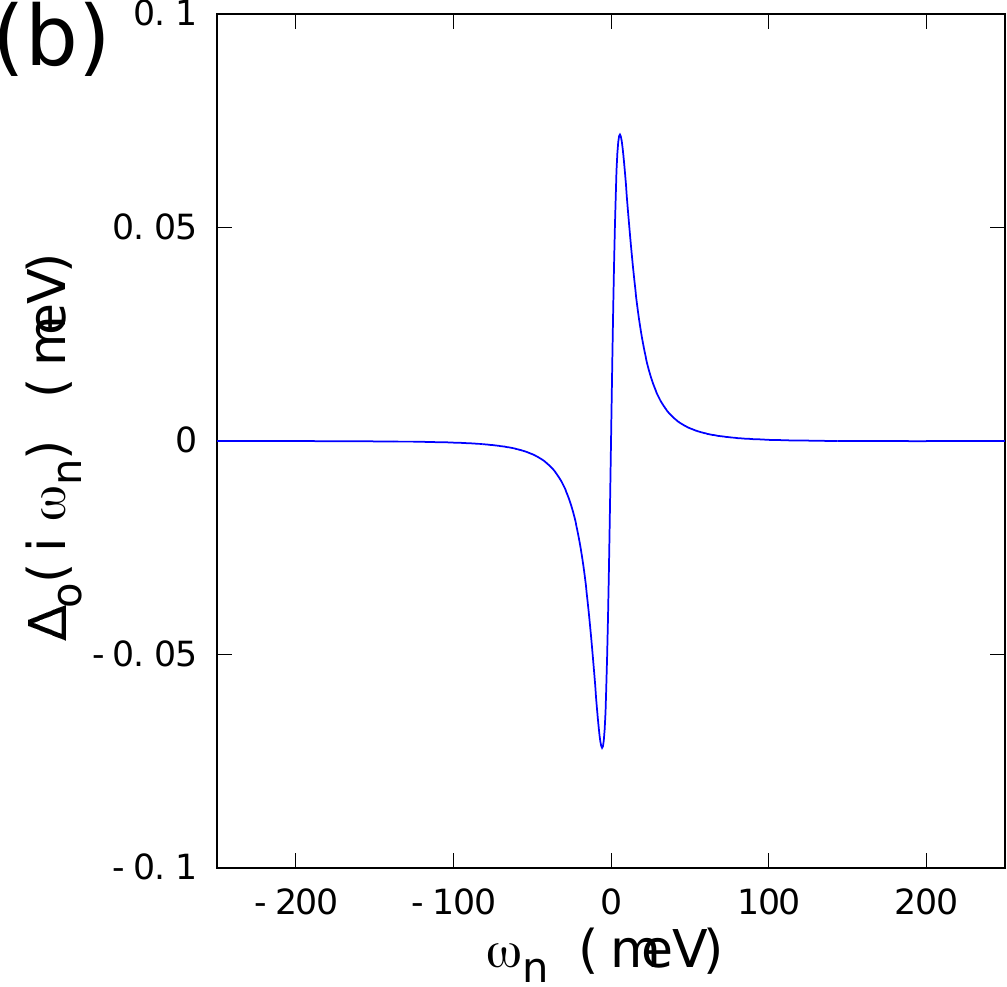}
\caption{Calculated  even $\Delta_{e}$ (left) and odd $\Delta_{o}$ (right)  frequency superconducting gap function for $\sr$ with $T$ = 1 K and magnetic field $h$ = 10 T.}\label{fig5}
\end{figure}

Having analyzed in detail the predicted signatures of strong coupling superconductivity in the tunneling spectra for the phosphides, we now focus on a different aspect of the non-trivial frequency dependence in strongly coupled superconductors, which is the possibility of magnetic field induced odd-frequency pairing \cite{Matsumoto2012,Aperis2015}. Using the same \textit{ab initio} input as previously, we solve the Eliashberg equations (\ref{eqS1})--(\ref{eqS3}) for different temperatures and magnetic fields. The Matsubara frequency dependence of typical self-consistent solutions for even ($\Delta_e(i\omega_n)$) and odd ($\Delta_e(i\omega_n)$) frequency superconductivity are shown in Fig.\ \ref{fig5}. Apart from the apparent difference in their overall shape, one can observe that for large frequencies $\Delta_e(i\omega_n)$ saturates to a negative value due to the effect of Coulomb repulsion. In contrast, $\Delta_o(i\omega_n)$ becomes zero at large frequencies. This is a manifestation of the ability of odd-frequency Cooper pairs to avoid completely Coulomb pair-breaking. This property is similar to the Coulomb avoidance of a sign-alternating, i.e., unconventional, superconducting gap in momentum space.

For even-frequency superconductivity $\Delta_e(i\omega_{n=1})$ coincides with the maximum gap and is therefore usually chosen as representative quantity for the order parameter. In the case of odd-frequency pairing, $\Delta_o(i\omega_{n=1})$ is small due to the oddness of the gap function and the odd gap acquires its maximum value at finite frequency. One can then select as an indicative order parameter the maximum value of the gap \cite{Matsumoto2012,Aperis2015}. On the other hand, it has been shown that the imaginary part of the zero real frequency component of the odd-frequency gap, $\rm{Im}\Delta_o(\omega=0)$), is non-zero and that this gives rise to in-gap states in the tunneling spectra  \cite{Aperis2015}. Therefore, we shall examine the (H,T)-dependence of both $\Delta_o(i\omega_{n=1})$ and $\rm{max}\Delta_o(i\omega_{n})$. 

\begin{figure*}[bth!]
\centering
\includegraphics[width=\linewidth]{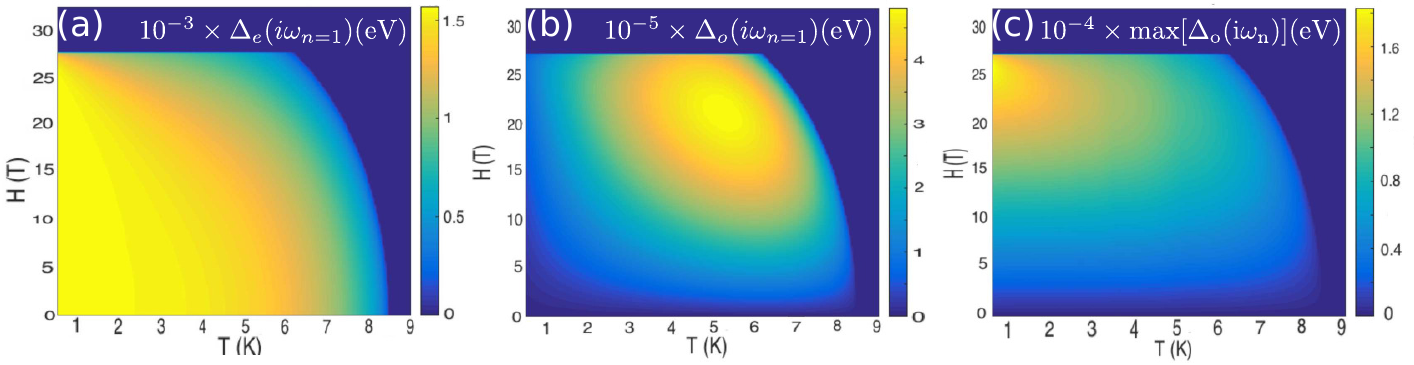}
\caption{Calculated magnetic field--temperature phase diagrams of $\sr$ for (a) the even-frequency gap function (b) odd-frequency gap function at $\omega_{n=1}$ and (c) maximum value of the odd-frequency gap function $\rm{max}\,\Delta_o(i\omega_n)$.} \label{fig6}
\end{figure*}

Our calculated (H,T) phase diagrams for $\Delta_e(i\omega_n)$ and $\Delta_o(i\omega_n)$ in the case of $\sr$ are shown in Fig.\ \ref{fig6}. It can be seen that the even-frequency gap exhibits the typical (H,T) phase diagram of a paramagnetic limited superconductor where, except from a narrow region near the critical field, it vanishes monotonically with temperature via a second order phase transition. As seen in Fig.\ \ref{fig6}(c) the maximum value of the odd frequency gap follows the same behavior, however, the first Matsubara  frequency odd-frequency gap exhibits a reentrant behavior with temperature as seen in Fig.\ \ref{fig6}(c). In addition, while $\Delta_e(i\omega_{n=1})$ decreases monotonically with the magnetic field,  both $\Delta_o(i\omega_{n=1})$ and $\rm{max}\,\Delta_o(i\omega_n)$ exhibit a reentrant behavior with the field. As we will show below, signatures of these markedly different behaviors may be identified in the superfluid density. Overall, the odd gap is always finite when the even gap is and one can see in Figs.\ \ref{fig6}(b),(c) that it is zero, too, in the absence of the magnetic field. In addition, its values are much smaller than those of the even-frequency gap. Thus, we see that the odd-frequency pairing is a subdominant order that is solely induced by the magnetic field, as discussed above and shown previously \cite{Matsumoto2012,Aperis2015}. Specifically, $\rm{max}\,\Delta_o(i\omega_{n})$ is one and $\Delta_o(i\omega_{n=1})$ is two orders of magnitude smaller than $\Delta_e(i\omega_{n=1})$. However, $\rm{max}\,\Delta_o(i\omega_{n})$ can be as large as 0.18\,meV.

\begin{figure*}[th!] 
\centering
\includegraphics[width=\linewidth]{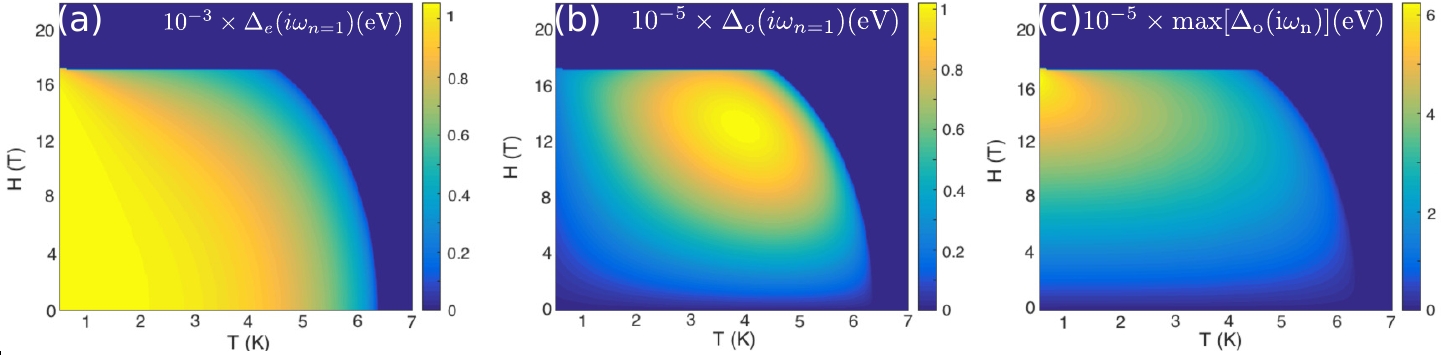}
\caption{Calculated magnetic field--temperature phase diagrams of $\ca$ for (a) the even-frequency gap function (b) odd-frequency gap function at $\omega_{n=1}$ and (c) maximum value of the odd-frequency gap function $\rm{max}\,\Delta_o(i\omega_n)$.} \label{fig7}
\end{figure*}

\begin{figure*}[th!] 
\centering
\includegraphics[width=\linewidth]{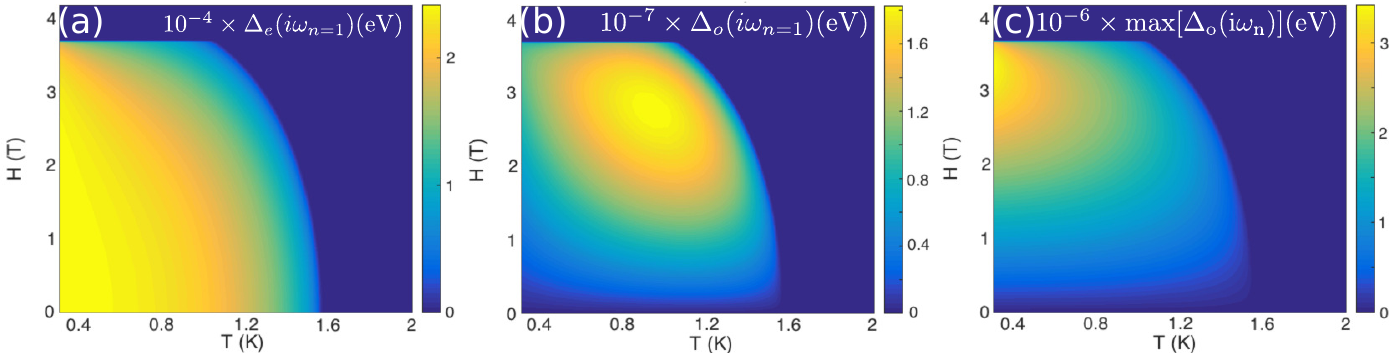}
\caption{Calculated magnetic field--temperature phase diagrams of $\la$ for (a) the even-frequency gap function, (b) odd-frequency gap function at $\omega_{n=1}$, and (c) maximum value of the odd-frequency gap function $\rm{max}\,\Delta_o(i\omega_n)$.} \label{fig8}
\end{figure*}

The calculated phase diagrams for $\ca$ and $\la$ are shown in Figs.\ \ref{fig7} and \ref{fig8}. They look very similar to those of $\sr$, but with the $T_c$'s and $H_c$'s becoming smaller as the coupling decreases from $\sr$ to $\la$. However, the relative magnitude between the even and odd superconducting components differs. For $\ca$, both $\rm{max}\,\Delta_o(i\omega_{n})$ and $\Delta_o(i\omega_{n=1})$ are two orders of magnitude smaller than $\Delta_e(i\omega_{n=1})$. For the weakly coupled $\la$, $\rm{max}\,\Delta_o(i\omega_{n})$ is two and $\Delta_o(i\omega_{n=1})$ is three orders of magnitude smaller than $\Delta_e(i\omega_{n=1})$. 
These findings suggest that as we move from a strongly retarded and coupled superconductor such as $\sr$ to the next family member compound, the fraction of Cooper pairs that can form odd-frequency pairing is heavily reduced. Therefore, we conclude that the effect of strong coupling  is crucial for the occurrence of odd-frequency superconductivity \cite{Kusunose2011a}.

Here it should be mentioned that the orbital depairing effect is not included in our theory. Given that this is a tremendously difficult task below $T_c$ while also including odd-frequency pairing \cite{Schossmann1986,Schossmann1986a}, we instead chose to confined ourselves to the inclusion of only the Zeeman effect. In any case, it is the latter effect that  causes the induced odd-frequency superconductivity. Naturally, this  approximation results in an overestimation of the upper critical field, e.g.\ by approximately $5-6$ times for $\sr$ \cite{Khasanov2014}. Therefore, below we will discuss penetration depth predictions as a function of the ratio $H/H_c$, with $H_c$ the critical field, rather than the actual field strength itself.

\begin{figure*}[th!]
\centering
\includegraphics[width=\linewidth]{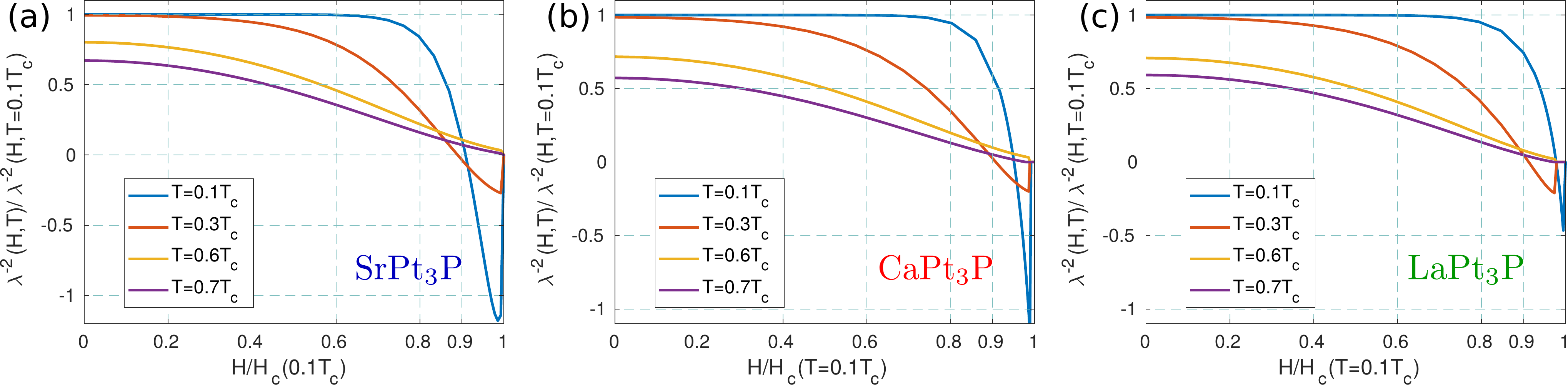}
\caption{The calculated normalized superfluid density as a function of the normalized magnetic field at several temperatures for (a) $\sr$, (b) $\ca$, and (c) $\la$. The negative regions exhibit a paramagnetic Meissner effect due to the magnetic field induced odd-frequency superconductivity.} 
\label{fig9}
\end{figure*}

\subsection{Experimental signatures of odd-frequency superconductivity: paramagnetic Meissner effect}

Inserting the results from the solution of Eqs.\ (\ref{eqS1})-(\ref{eqS3}) into Eq.\  (\ref{londonfinal}), we calculated the normalized superfluid density, shown in Fig.\ \ref{fig9} for all three compounds. Interestingly, at low temperatures and close to the critical field, a region occurs where the superfluid density becomes negative. This change in sign signals a change in the Meissner response of the superconductor from diamagnetic to paramagnetic. As the temperature is increased, the paramagnetic region is lost and the diamagnetic Meissner effect is recovered. Comparing with Figures \ref{fig6}, \ref{fig7}, and \ref{fig8} we observe that the occurrence of the paramagnetic Meissner effect follows the (H,T) dependence of the odd-frequency superconducting component, $\rm{max}\,\Delta_o(i\omega_n)$, and therefore, the experimental observation of our predicted field dependence of the superfluid density  could verify the existence of such pairing. As expected, the region of magnetic field values where this effect takes place expands as we move from weak to strong coupling. This trend could also serve as a signature for the evolution of the odd-frequency gap strength across the APt$_3$P series.

 \begin{center}
  \begin{ruledtabular}
  \begin{table*}[tbh!]
 \caption{Frequency dependent phenomena in relation to the strong to weak coupling phenomenology in the APt$_3$P superconductors. For each material are shown the coupling strength $\lambda$, the logarithmic phonon frequency $\omega_{\text{ln}}$, the transition temperature $T_{c}$, the ratios $2\Delta(0)/T_{c}$ and $T_{c}/\omega_{\rm ln}$, and, additionally, the ratio of the energy width of the dip-hump region over $\Delta(0)$, the ratio between the maximum magnetic field induced odd-frequency superconducting gap over it's even-frequency counterpart, and the relative difference between the paramagnetic critical field and the field where the paramagnetic Meissner effect sets in. All values were calculated based on the \textit{ab initio} input from Ref.\ \cite{Subedi2013}.}
 \label{table2}
 \begin{tabular}{c |c c c c c c c c c c}
  $\phantom{a}$&$\lambda$& $\omega_{\text{ln}}$ (K) & $T_{c}$ (K) & $2\Delta(0)/T_{c}$ & $T_{c}/\omega_{\ln}$ & $\phantom{a}$&\Large $\frac{\omega_{hump}-\omega_{dip}}{\Delta(0)}$\normalsize & \Large$\frac{\rm{max}\Delta_o(i\omega_n)}{\Delta_e(i\omega_{n=1})}\normalsize$ & \Large$\frac{\rm{H}_c-\rm{H}_{par}}{\rm{H}_c}|\normalsize^{\phantom{a}}_{T=0.1T_c}\normalsize$\\
 \hline \\[-0.2cm]
 SrPt$_{3}$P & 1.33 & 77 & 8.56 & 4.24 & 0.111 & & 0.119 & 0.107 & 0.094\\
 CaPt$_{3}$P & 0.85 & 110 & 6.45 & 3.79 & 0.058 & & 0.080 & 0.063 & 0.043\\
 LaPt$_{3}$P & 0.57 & 118 & 1.58 & 3.59 & 0.013 & & 0.022 & 0.017 & 0.015\\
 \end{tabular}
 \end{table*}
 \end{ruledtabular}
 \end{center}

As a closing remark, we collect our findings from the previous sections regarding the location of the dip-hump structures in the tunneling, the strength of the magnetic field induced odd-frequency superconducting gap and the field region where the concomitant paramagnetic Meissner effect appears as the coupling strength varies from strong in $\sr$ to weak in $\la$ in Table \ref{table2}. By dividing the energy width of the dip-hump structure with the respective zero temperature, zero field (even-frequency) superconducting gap, we add column six of the table. Furthermore, by dividing the  maximum calculated odd-frequency induced superconducting gap by the respective even frequency one at zero field and temperature we fill column seven. Lastly, we fill the eighth column by the calculated relative difference between the magnetic field field value where the paramagnetic Meissner effect first appears, $H_{par}$, with the critical magnetic field where superconductivity is destroyed. Remarkably, we observe that the values for each of the APt$_3$P family member is in impressive agreement with their respective $T_c/\omega_{ln}$ ratio. This finding provides a direct link between seemingly unrelated experimentally measurable quantities such as the zero field, low temperature tunneling spectra and the high field, low temperature paramagnetic Meissner effect in this family of superconductors. Even more interestingly, it connects the outcome of a zero magnetic field tunneling experiment with the formation of odd-frequency superconductivity at high magnetic fields. Whether the above trends are a peculiarity of the phosphides series or may have more a general application remains as an intriguing question for future investigation.

\section{Conclusions}

As one traverses the APt$_3$P series of superconductors retardation and coupling strength vary from weak to strong, as is evidenced by the increasing $2\Delta(0)/T_c$ and $T_c/\omega_{ln}$ ratios in Table \ref{table1}. This property of this  phosphides series provides a unique opportunity for testing the manifestations of strong-coupling phenomena experimentally in a relatively controlled manner. Two such phenomena, that seem otherwise unrelated to each other, are the appearance of fine structures in the superconducting tunneling spectra known as dip-humps and the anticipated formation of odd-frequency superconductivity by applied magnetic fields.
In this work, we presented detailed Eliashberg theory calculations for these phenomena across the APt$_3$P (A = $\sr$, $\ca$, and $\la$) compounds using as input electron, phonon and electron-phonon properties of these materials that were calculated by first principles. 

Our predictions could  serve as a multi-check experimental protocol for the ultimate detection of odd-frequency superconductivity in this family of superconductors. Thus, they pave a  way towards the unambiguous observation of this so far elusive state in a bulk superconductor \cite{Linder2017}. Moreover, such a protocol could be applied to other groups of superconducting materials that share similar coupling and retardation properties where experimental conditions for the observation of the paramagnetic Meissner effect could be further optimized. Promising examples could be the recently proposed two-dimensional single and few-layer MgB$_2$ superconductors \cite{Bekaert2017,Bekaert2019,Aperis2015}.

\acknowledgments
This paper is dedicated to Professor Gerasim Eliashberg on the occasion of his 90$^{\rm th}$ birthday.
We gratefully acknowledge support from the Swedish Research Council (VR) and from the Swedish National Infrastructure for Computing (SNIC). We acknowledge fruitful discussions with F.\ Schrodi and G.\ Varelogiannis.

\bibliographystyle{apsrev4-1}
%

\end{document}